\documentclass{article}
\usepackage{spconf,amsmath,epsfig}
\usepackage{xcolor}
\usepackage{amssymb}
\usepackage{algorithm,algcompatible,amsmath}
\usepackage{multirow}
\usepackage{siunitx}
\usepackage{booktabs}
\usepackage{algorithm} 
\usepackage{color, colortbl}
\usepackage{enumitem}
\usepackage{algpseudocode} 
\usepackage{amsthm}
\usepackage{comment}
\usepackage{cuted}
\newtheorem{proposition}{Proposition}

\newtheorem{definition}{Definition}
\newtheorem{lemma}{Lemma}
\let\OLDthebibliography\thebibliography
\renewcommand\thebibliography[1]{
  \OLDthebibliography{#1}
  \setlength{\parskip}{0pt}
  \setlength{\itemsep}{0pt plus 0.3ex}
}
\usepackage{multicol}

\ninept
\begin{document}\sloppy

\def\bb{\mathbb}
\def\cal{\mathcal}
\def\bsym{\boldsymbol}
\def\tr{\textnormal{tr}}
\def\blktr{\textnormal{blktr}}
\def\i{\textnormal{i}}
\def\o{\textnormal{o}}
\def\int{\textnormal{in}}
\def\out{\textnormal{out}}
\def\blf{\mathbf}
\title{Graph filtering over expanding graphs}
%
\name{Bishwadeep Das, Elvin Isufi}
\address{B.Das@tudelft.nl, E.Isufi-1@tudelft.nl}

\maketitle

\begin{abstract}
Our capacity to learn representations from data is related to our ability to design filters that can leverage their coupling with the underlying domain. Graph filters are one such tool for network data and have been used in a myriad of applications. But graph filters work only with a fixed number of nodes despite the expanding nature of practical networks. Learning filters in this setting is challenging not only because of the increased dimensions but also because the connectivity is known only up to an attachment model. We propose a filter learning scheme for data over expanding graphs by relying only on such a model. By characterizing the filter stochastically, we develop an empirical risk minimization framework inspired by multi-kernel learning to balance the information inflow and outflow at the incoming nodes. We particularize the approach for denoising and semi-supervised learning (SSL) over expanding graphs and show near-optimal performance compared with baselines relying on the exact topology. For SSL, the proposed scheme uses the incoming node information to improve the task on the existing ones. These findings lay the foundation for learning representations over expanding graphs by relying only on the stochastic connectivity model.
\end{abstract}
\section{Introduction}
Graph filters are parametric information processing architectures for network data \cite{shuman_emerging_2013,ortega_graph_2018} with wide applicability in signal de-noising \cite{chen2014signal,hua2019learning}, recommender systems \cite{huang_rating_2018,isufi2021accuracy}, semi-supervised learning \cite{sandryhaila_discrete_2013,berberidis2018adaptive}, and graph-based dictionary representations \cite{thanou2014learning}. By relying on information exchange between neighbouring nodes, graph filters extend the convolution operation to the graph domain \cite{sandryhaila_discrete_2013,gama2020graphs}. In turn, by learning the convolution parameters, they can account for the network data-topology coupling to solve the task at hand. However, graph filters are used only for graphs with a fixed number of nodes despite the evidence that practical graphs often grow in size \cite{erdos_evolution_1961,barabasi_emergence_1999,bianconi_competition_2001}. Filtering network data in this setting is challenging not only because of the increase in graph dimension but also because we do not know how the new nodes attach to the graph.

The importance of processing data over expanding graphs and the challenges arising when learning a filter on them have been recently recognized in a few works. Authors in \cite{chen_semi-supervised_2014} focus on semi-supervised learning with incoming nodes. First, a filter is learned to solve the task on the existing nodes, and then the filter output is used as a feature vector to predict the label of a new incoming data-point node. The work in \cite{cervino2021increase} learns a graph filter-based neural network over a sequence of growing graphs, which are generated from a common graphon model \cite{lovasz2012large}. However, the generated graphs are not related to each other. The works in \cite{venkitaraman_recursive_2020,shen_online_2019} perform regression tasks over graphs but rely on the attachment of the incoming nodes.

Despite showing potential, these works rely on the exact topological connectivity or available node features. However, in many cases, we only have the stochastic attachment model for this growing graph. Hence, deploying graph filters in this setting leads to a stochastic output, which requires a statistical approach in the filter learning phase. To fill this gap, we propose a stochastic graph filtering framework over an expanding graph without knowing the connectivity of the incoming nodes. Our detailed contribution with this framework is threefold:


\begin{enumerate}[label=\textbf{C.\arabic*.}, start = 1]
\item We propose a filtering model over expanding graphs that relies only on the preferential attachment information. The model comprises two parallel graph filters: one operating on a graph in which directed edges land at the incoming node; and one operating on an expanded graph in which directed edges depart from the incoming node. Such a procedure allows more flexibility to control the information inflow and outflow on the expanded graph and greater mathematical tractability compared with a filter running over a single graph.
\item We adapt the proposed approach to signal de-noising and semi-supervised learning over expanded graphs. We characterise the filter output stochastically and show the role played by the filter parameters and the attachment model.
\item We develop an empirical risk minimization framework to learn the filters. This framework is inspired by multi-kernel learning and balances the information between the two graphs via a single parameter.
\end{enumerate}
\noindent Numerical results with synthetic and real data from sensor and blog networks corroborate our findings.

\vspace{-5mm}
\section{Problem Formulation}\label{Section PF}
Consider a graph $\mathcal{G}=(\mathcal{V},\mathcal{E})$ with node set $\mathcal{V} = \{v_1, \ldots, v_N\}$, edge set $\mathcal{E}$, and adjacency matrix $\mathbf{A}$. An incoming node $v_+$ attaches to $\mathcal{G}$ and forms two sets of directed edges: a set $\{( v_n,v_+)\}$ starting from $v_+$ and landing at the existing nodes $v_n$, whose weights are collected in vector $\mathbf{b}^\i_+\in \mathbb{R}^N$; and another set of $\{(v_+,v_n)\}$ starting from the existing nodes $v_n\in\cal V$ and landing at $v_+$, whose weights are collected in vector $\mathbf{a}^\o_+\in \mathbb{R}^N$. We represent these connections with two directed graphs $\mathcal{G}^{\i}_{+} = \big(\{\mathcal{V} \cup v_+\}, \{\mathcal{E} \cup ( v_n,v_+)\}\big)$ and $\mathcal{G}^{\o}_{+} = \big(\{\mathcal{V} \cup v_+\}, \{\mathcal{E} \cup (v_+,v_n)\}\big)$ whose adjacency matrices are
    \begin{equation}\label{jock}
        \mathbf{A}^{\i}_{+}=\begin{bmatrix}
        \mathbf{A} & \mathbf{b}^\i_{+} \\
        \mathbf{0}^{\top} & 0 \\\end{bmatrix}~~~\textnormal{and}~~~
\mathbf{A}^\o_{+}=\begin{bmatrix}
        \mathbf{A} & \mathbf{0} \\
        \mathbf{a}^{\o\top}_{+} & 0 \\ \end{bmatrix}
    \end{equation}
respectively and where $^\top$ denotes the transpose and $\mathbf{0}$ the all-zero vector. A conventional way to model the attachment of incoming nodes is via stochastic models \cite{barabasi2016network} in which node $v_+$ connects to $v_l$ with probability $p^{\i}_l$ and weight $w_l^{\i}$ in graph $\mathcal{G}^{\i}_{+}$, and probability $p^{\o}_i$ and weight $w_l^{\o}$ in graph $\mathcal{G}^{\o}_{+}$. Hence, $\mathbf{b}^\i_+$ and $\mathbf{a}^\o_+$ are random vectors with expected values $\bsym \mu^{\i}\!=\!\mathbf{w}^{\i}\!\circ\!\mathbf{p}^{\i}$ and  $\bsym \mu^{\o}\!=\!\mathbf{w}^{\o}\!\circ\!\mathbf{p}^{\o}$, and covariance matrices $\bsym{\Sigma}^\i$ and $\bsym{\Sigma}^\o$, respectively. Here we define $\mathbf{w}^{\i}\!=\![w_1^{\i},\!\ldots,\!w_N^{\i}]^{\top}$, $\mathbf{w}^{\o}\!=\![w_1^{\o},\!\ldots,\!w_N^{\o}]^{\top}$ and denote by $\circ$ the Hadamard product.

While analyzing expanding graphs is an important topic, in this paper we are interested in processing data defined over the nodes of these graphs. Let then $\blf{x}_+=[\blf{x}^\top,x_+]^{\top} \in \mathbb{R}^{N+1}$ be a set of signal values over nodes $\mathcal{V} \cup v_+$ in which vector $\mathbf{x}=[x_1,\ldots,x_N]^{\top } \in \mathbb{R}^{N}$ collects the signals for the existing nodes $\mathcal{V}$ and $x_+$ is the signal at the incoming node $v_+$. Processing signal $\blf{x}_+$ amounts to designing graph filters that can capture its coupling w.r.t. the underlying directed graphs $\mathcal{G}^{\i}_{+}$ and $\mathcal{G}^{\o}_{+}$. To do so, we consider a filter bank of two convolutional filters \cite{ortega_graph_2018,sandryhaila_discrete_2013}, one operating on graph $\mathcal{G}^{\i}_{+}$ and one on graph $\mathcal{G}^{\o}_{+}$. Mathematically, with $\mathbf{h}^{\i} = [h_0^{\i}, \ldots, h_L^{\i}]^\top$ and $\mathbf{h}^{\o} = [h_0^{\o}, \ldots, h_M^{\o}]^\top$ representing the vector of coefficients for filters $\mathbf{H}^\i(\mathbf{A}^{\i}_{+})$ and $\mathbf{H}^\o_{+}(\mathbf{A}^\o_{+})$, respectively, the filter bank output is
    \begin{equation}\label{ncr}
    \mathbf{y}_+=\blf y^{\i}_+ + \blf {y}^\o_+:=\underbrace{\sum_{l=0}^L h^\i_l[\mathbf{A}_+^{\i}]^{l}}_{\mathbf{H}^\i(\mathbf{A}^{\i}_{+})}\mathbf{x}_+ +\underbrace{\sum_{m=0}^M h^\o_m[\blf A_+^{\o}]^{m}}_{\mathbf{H}^\o_{+}(\mathbf{A}^\o_{+})}\mathbf{x}_+
    \end{equation}
where without loss of generality we consider $\mathbf{y}_+ = [\mathbf{y}^\top, y_+]^\top$ and $\blf y^\i_+$ and $\blf y^\o_+$ are the outputs over graphs $\mathcal{G}^{\i}_{+}$ and $\mathcal{G}^{\o}_{+}$, respectively.

The stochastic nature of the attachment yields a random $\mathbf{y}_+$. It is also notoriously challenging to compute statistical moments of powers of the adjacency matrices of increasing graphs because third or higher-order moments of the attachment pattern of the incoming node appear\footnote{This is a challenge if we consider a single graph $\mathcal{G}_+ = (\mathcal{V}_+, \mathcal{E}_+)$ whose edge set contains both the incoming and outgoing edges at node $v_+$.} \cite{dai2013multivariate}. However, because of the decoupled nature between the incoming and outgoing edges at node $v_+$, the $k$th powers of the adjacency matrices have the block structure
     \begin{equation}\label{eq.blockPower}
        [\mathbf{A}^{\i}_{+}]^k=\begin{bmatrix}
        \mathbf{A}^k & \mathbf{A}^{k-1}\mathbf{b}^\i_{+} \\
        \mathbf{0}^{\top} & 0 \\\end{bmatrix}~~\textnormal{and}~~[\mathbf{A}^\o_{+}]^k=\begin{bmatrix}
        \mathbf{A}^k & \mathbf{0} \\
        \mathbf{a}^{\o\top}_{+}\mathbf{A}^{k-1} & 0 \\ \end{bmatrix}
    \end{equation}
which facilitate the stochastic analysis of the filter output [cf. Sec.~\ref{Section Methods}].


Given the filter bank in \eqref{ncr}, our goal translates into estimating the filter coefficients $\mathbf{h}^\i$ and $\mathbf{h}^\o$ to solve specific learning tasks in a statistical fashion \cite{vapnik1999overview}. Specifically, we consider a training set $\mathcal{T}=\{(v_{+},\mathbf{x}_{+},\mathbf{t}_{+})\}$ comprising a set of incoming nodes $v_+$ w.r.t. a fixed existing graph $\mathcal{G}$, an expanded graph signal $\mathbf{x}_{+}$ (e.g., noisy or partial observations), and a target output signal $\mathbf{t}_{+}=[\mathbf{t}^\top,t_{+}]^{\top}$ (e.g., true signal, or class labels). Then, we learn the filter by solving
\begin{align}\label{pbasic}
\begin{split}
    \underset{\mathbf{h}^\i, \mathbf{h}^\o}{\text{min }}\hspace{1mm} \frac{1}{2 \gamma}\mathbb{E}\big[f_{\mathcal{T}}(\mathbf{h}^{\i},\mathbf{h}^{\o},\mathbf{t}_+)\big]& +\frac{1}{2\alpha}g(\mathbf{h}^\i)+\frac{1}{2(1-\alpha)}j(\mathbf{h}^\o)\\
\end{split}
\end{align}
where $\bb E[f_{\mathcal{T}}(\mathbf{h}^{\i},\mathbf{h}^{\o},\mathbf{t}_+)]$ is the expected task-specific loss with the expectation taken w.r.t. both the graph attachment vectors $\mathbf{b}^\i_+$, $\mathbf{a}^\o_+$ and the data distribution; and $g(\cdot)$, $j(\cdot)$ are filter-specific regularizers (e.g., norm two of coefficient vectors) to avoid overfitting. The regularization weight $\gamma >0$ controls the trade-off between fitting and regularization and scalar $0 < \alpha < 1$ balances the impact between the two filters inspired by multi-kernel learning \cite{rakotomamonjy2008simplemkl}.

In the next section, we shall particularize problem \eqref{pbasic} to graph signal de-noising (Sec.~\ref{subsec_den}) and graph-based semi-supervised learning (Sec. \ref{subsec_SSL}). For both settings, we shall characterize the filter output stochastically, use its first- and second-order moments in \eqref{pbasic}, and show the role played by the attachment models on $\mathcal{G}^{\i}_{+}$ and $\mathcal{G}^{\o}_{+}$.


\vspace{-5mm}
\section{Filtering with Incoming Nodes }\label{Section Methods}

Before defining the filter bank in \eqref{ncr} for the two tasks, we first rearrange it in a compact form, instrumental for our analysis. This form will also show the influence of the incoming node connectivity on the filter output.

\subsection{Compact Form}

The goal of this section is to isolate the filter coefficients $\mathbf{h} = [\mathbf{h}^{\i, \top}, \mathbf{h}^{\o, \top}]^\top$ and write \eqref{ncr} as $\mathbf{y}_+=\mathbf{W}_+\mathbf{h}$, where $\mathbf{W}_+$ contains the coupling between the stochastic expanded graphs and the signal.

Analyzing filter $\mathbf{H}^\i(\mathbf{A}^{\i}_{+})$, it is possible to write its output as
\begin{equation}\label{eq.out_incoming}
    \mathbf{y}_+^{\i} = [\mathbf{x}_+, \mathbf{A}^{\i}_{+}\mathbf{x}_+, \ldots,[\mathbf{A}^{\i}_{+}]^L\mathbf{x}_+]\mathbf{h}^{\i}.
\end{equation}
Then, leveraging the structure of the input $\mathbf{x}_+ = [\mathbf{x}^{\top}, x_+]^\top$ and the block-structure of $[\mathbf{A}^{\i}_{+}]^k$ in \eqref{eq.blockPower}, we can write \eqref{eq.out_incoming} as
\begin{equation}\label{eq.outInc}
    \mathbf{y}_+^{\i} =\begin{bmatrix}
        \widehat{\mathbf{L}}_x\\
        \mathbf{x}_L^{\top}\\
        \end{bmatrix}
        \mathbf{h}^\i = 
        \begin{bmatrix}
        \mathbf{L}_x+x_+\overline{\mathbf{L}}_b\\
        \mathbf{x}_L^{\top}\\
        \end{bmatrix}
        \mathbf{h}^\i
\end{equation}
where $\widehat{\mathbf{L}}_x = \mathbf{L}_x+x_+\overline{\mathbf{L}}_b
$, $\mathbf{L}_x=[\mathbf{x},\mathbf{Ax},\ldots,\mathbf{A}^L\mathbf{x}]$ and $\overline{\mathbf{L}}_b=[\mathbf{0},\mathbf{b}^\i_{+},\ldots,\mathbf{A}^{L-1}\mathbf{b}^\i_{+}]$ are $N\times(L+1)$ matrices and $\mathbf{x}_L=[x_+,0,\ldots,0]^{\top}\in\bb R^{L+1}$. Eq. \eqref{eq.outInc} shows that the output $\mathbf{y}^{\i}$ on the existing nodes nodes $\mathcal{V}$ is influenced by propagating their own signal $\mathbf{x}$ [cf. $\mathbf{L}_x$] and by propagating signal $x_+$ of $v_+$ w.r.t. the incoming attachments $\mathbf{b}_+$ [cf. $\overline{\mathbf{L}}_b$]. Instead, the filter output at the incoming node $y_+^{\i}$ is just a scaled version of the input $x_+$ by coefficient $h_0^{\i}$. The latter is because edges on graph $\mathcal{G}^{\i}$ leave node $v_+$ and land on $\mathcal{V}$; hence, governing the direction of the signal propagation.

Likewise, analyzing filter $\mathbf{H}^\o(\mathbf{A}^{\o}_{+})$, we can write its output as
\begin{equation}\label{eq.out_outgoing}
    \mathbf{y}_+^{\o} = [\mathbf{x}_+, \mathbf{A}^{\o}_{+}\mathbf{x}_+, \ldots,[\mathbf{A}^{\o}_{+}]^M\mathbf{x}_+]\mathbf{h}^{\o}.
\end{equation}
Leveraging again the structure of the input and that of the matrix powers $[\mathbf{A}^{\o}_{+}]^k$ in \eqref{eq.blockPower}, allows writing \eqref{eq.out_outgoing} as
    \begin{equation}\label{eq.outOut}
        \mathbf{y}_+^{\o}= 
        \begin{bmatrix}
        \mathbf{M}_x\\
        \widehat{\mathbf{m}}_{x}^{\top}\\
        \end{bmatrix}
        \mathbf{h}^\o
        =\begin{bmatrix}
        \mathbf{M}_x\\
        \mathbf{a}_+^{\o\top}\overline{\mathbf{M}}_{x}+\mathbf{x}_M^{\top}\\
        \end{bmatrix}
        \mathbf{h}^\o.
    \end{equation}
where $\mathbf{M}_x = [\mathbf{x},\mathbf{Ax},\ldots,\mathbf{A}^M\mathbf{x}]$ is an $N \times (M+1)$ matrix, $\widehat{\mathbf{m}}_{x} = \overline{\mathbf{M}}_{x}^{\top}\mathbf{a}_+^{\o}+\mathbf{x}_M$, $\overline{\mathbf{M}}_{x} = [\mathbf{0},\mathbf{x},\ldots,\mathbf{A}^{M-1}\mathbf{x}]$ are of dimensions $(M+1)\times 1$ and $N\times (M+1)$, respectively, and $\mathbf{x}_M=[x_+,0,\ldots,0]^{\top}\in\bb R^{M+1}$. That is, the output $\mathbf{y}^{\o}$ on the existing nodes $\mathcal{V}$ is influenced only by propagating their own signal $\mathbf{x}$. Instead, the output $y_+^{\o}$ at node $v_+$ comprises: (i) the match between the attachment pattern $\mathbf{a}_+^{\o\top}$ and the signal shifted over the existing graph $\mathcal{G}$, $\overline{\mathbf{M}}_{x}$; i.e., $\mathbf{a}_+^{\o\top}\overline{\mathbf{M}}_{x}$; and (ii) a scaled version of its own signal $x_+$ by coefficient $h_0^{\o}$. The output on the existing nodes $\mathcal{V}$ is not influenced by signal $x_+$ because the edges in $\mathcal{G}^{\o}$ leave those nodes and land on $v_+$. The latter is also justified by the structure of matrix $\overline{\mathbf{M}}_{x}$; i.e., the existing signal $\mathbf{x}$ is first percolated over $\mathcal{G}$ and then mapped onto $v_+$ through its attachment pattern $\mathbf{a}_+$ in a \emph{matched filtering} principle \cite{turin1960introduction}.

Bringing together \eqref{eq.outInc} and \eqref{eq.outOut}, leads to the desired compact form
\begin{equation}\label{eq.filt_compact}
    \mathbf{y}_+=\blf y^{\i}_+ + \blf {y}^\o_+
        = \mathbf{W}_+\mathbf{h}~~~\text{and}~~~\mathbf{W}_+ = \begin{bmatrix}
        \widehat{\mathbf{L}}_x &\mathbf{M}_x\\
        \mathbf{x}_L^{\top} & \widehat{\mathbf{m}}_{x}^{\top}
        \end{bmatrix}.
\end{equation}

\smallskip
\noindent\textbf{Statistical identity.} Throughout the statistical analysis of the filter output $\mathbf{y}_+$, we will deal with expectations of the form $\bb E [\blf{L}_x^{\top}\blf{C}\blf{M}_x]$ for some $N \times N$ square matrix $\blf{C}$. In the remainder of this section, we derive a handy formulation for it by using the compact form \eqref{eq.filt_compact}. For this, we will need the block-trace operator as defined next.

\begin{definition}[Block trace]\label{hmm}
\textit{Let $\mathbf{Z}$ be a block matrix comprising $N\times N$ sub-matrices $\mathbf{Z}_{ij}$. The block trace operator takes as arguments matrix $\mathbf{Z}$ and an $N \times N$ matrix $\mathbf{Y}$ to yield a matrix $\mathbf{U}=\blktr(\mathbf{Z},\mathbf{Y})$ with $(i,j)$ entry $U_{ij}=\text{\tr}(\mathbf{Y}\mathbf{Z}_{ij})$ and $\text{\tr}(\cdot)$ being the trace operator.}
\end{definition}
%

\begin{lemma}\label{velvet}
\textit{
Given an existing graph $\cal G$ with a noisy graph signal $\mathbf{x} = \mathbf{t} + \mathbf{n}$, where $\mathbf{t}$ is the true signal and $\mathbf{n}$ a Gaussian noise $\mathcal{N}(\mathbf{0}, \sigma^2\mathbf{I})$. Consider also matrices $\blf{L}_x$ and $\blf{M}_x$ [cf. \eqref{eq.outInc} and \eqref{eq.outOut}], which can be expanded further as
\begin{equation}\label{house}
      \blf{L}_x=\blf{L}(\mathbf{I}_{L+1}\otimes\mathbf{x})~~~\textnormal{and}~~~\blf{M}_x=\blf{M}(\mathbf{I}_{M+1}\otimes\mathbf{x})
\end{equation}
where $\blf{L} = [\mathbf{I},\mathbf{A},\ldots,\mathbf{A}^{L}]$, $\blf{M} = [\mathbf{I},\mathbf{A},\ldots,\mathbf{A}^{M}]$, $\mathbf{I}_N$ is the $N\times N$ identity matrix, and $\otimes$ is the Kronecker product. Then, for any $N\times N$ square matrix $\mathbf{C}$ the following identity holds:
\begin{equation}
\bb E [\blf{L}_x^{\top}\blf{C}\blf{M}_x] = \blf{L}_t^{\top} \blf{C}\blf{M}_t +\sigma^2\blktr(\blf{L}^\top\blf{C}\blf{M},\mathbf{I}_N)
\end{equation}
where $\blf L_t=\blf L_x|_{x=t}$, $\blf M_t=\blf M_x|_{x=t}$, and $\blktr(\cdot)$ is the block operator in Def. \ref{hmm}.
}
\end{lemma}

\noindent\textit{Proof.}
See the appendix in the Supplementary Material.\hfill\qed


\subsection{Signal Denoising}\label{subsec_den}

The first task we are interested in is recovering a true signal $\mathbf{t}_+$ from its noisy observations $\mathbf{x}_+$ by knowing only the stochastic attachment pattern of the incoming node. For this, we consider as cost the mean squared error $\mathbb{E}\big[f_{\mathcal{T}}(\mathbf{h},\mathbf{t}_+)\big]:=\mathbb{E}[||\mathbf{W}_+\blf h-\mathbf{t}_+||_\mathbf{D}^2]$, where $\mathbf{D}=\textnormal{diag}(d_1,\ldots,d_{N+1})\in\{0,1\}^{N+1\times N+1}$ is a diagonal matrix with $d_n=1$ only if account for the MSE at node $n$ and zero otherwise. The following proposition quantifies the latter.


\begin{proposition}\label{zion}
\textit{Given a graph $\mathcal{G} = (\mathcal{V}, \mathcal{E})$ with adjacency matrix $\mathbf{A}$ and an incoming node $v_+$ connecting to $\mathcal{G}$ with random attachment vectors $\bf b^\i$ and $\bf a^\o$ with respective means $\bsym{\mu}^\i$, $\bsym{\mu}^\o$ and covariance matrices $\bsym{\Sigma}^\i$, $\bsym{\Sigma}^\o$ [cf. \eqref{jock}]. Consider a noisy signal $\mathbf{x}_+=\mathbf{t}_++\mathbf{n}_+$ over the nodes $\mathcal{V} \cup v_+$, with $\mathbf{x}_+ = [\mathbf{x}^\top, x_+]$, $\mathbf{t}_+ = [\mathbf{t}^\top, t_+]$, and $\mathbf{n}_+ \sim \mathcal{N}(\mathbf{0}, \sigma^2\mathbf{I}_{N+1})$. Let also $\bf y_+=\bf W_+\bf h$ $[cf. \eqref{eq.filt_compact}]$ be the filtered output. Then, the MSE of the filter output $\textnormal{MSE}_{\mathbf{D}}(\mathbf{h}) = \mathbb{E}[||\mathbf{W}_+\blf h-\mathbf{t}_+||_\mathbf{D}^2]$ computed on a set of nodes sampled by the diagonal matrix $\mathbf{D}=\textnormal{diag}(d_1,\ldots,d_{N+1})\in\{0,1\}^{N+1\times N+1}$ is
}
%
%
%
    \begin{equation}\label{sure}
      \textnormal{MSE}_{\mathbf{D}}(\mathbf{h})=\blf h^{\top}\boldsymbol{\Delta}\blf h-2\blf h^{\top}\boldsymbol{\theta}+||\mathbf{t}_+||_{\mathbf{D}}^2  
    \end{equation}
\textit{where} $\boldsymbol{\Delta}=[\bsym{\Delta}_{11},\bsym{\Delta}_{12};\bsym{\Delta}_{21},\bsym{\Delta}_{22}]$ is a $2 \times 2$ block matrix with:
%
    %
\begin{align}\label{eq.Delta11}
\begin{split}
&\boldsymbol{\Delta}_{11}\!=\!\blf L_t^{\top}\blf{D}_N\blf L_t \!+\!\sigma^2\blktr(\blf L^{\top}\blf{D}_N\blf L,\blf I_N)+t_+\blf L_t^{\top}\blf D\overline{\mathbf{L}}_{\mu^\i}\\
 &\quad+t_+\overline{\mathbf{L}}^{\top}_{\mu^\i}\blf{D}_N\blf L_t+(t_+^2+\sigma^2)(\overline{\mathbf{L}}^{\top}_{\mu^\i}\blf D\overline{\mathbf{L}}_{\mu^\i}\\
 &\quad+\blktr(\overline{\mathbf{L}}^{\top}\blf D\overline{\mathbf{L}},\boldsymbol{\Sigma}^{\i}))+d_{N+1}\textnormal{diag}(t_+^2+\sigma^2,\blf{0}_L)
\end{split}
\end{align}
where $\blf D=\textnormal{diag}(d_1,\ldots,d_N)$ contains the indices of the sampled nodes in $\cal G$, $\blf L=[\blf I,\ldots,\blf A^L]$, $\blf L_t=[\blf t, \blf A\blf t,\ldots,\blf A^L\blf t]$, $\overline{\blf L}\!=\![\blf 0,\blf I,\blf A, \ldots, \blf A^{L\!-\!1}]$, \!\!$\overline{\blf L}_t\!=\![\blf 0,\blf t,\blf A\blf t, \ldots, \blf A^{L\!-\!1}\blf t]$, and $\overline{\blf L}_{\mu^\i} = \overline{\blf L}_t|_{t = \mu^\i}$
%
\begin{align}\label{eq.Delta12}
\begin{split}
&\boldsymbol{\Delta}_{12} = \boldsymbol{\Delta}_{12}^{\top}=\blf L_t^{\top}\blf D\blf M_t+t_+\overline{\mathbf{L}}_{\mu^\i}\blf D\blf M_t\\
&+\sigma^2\blktr(\blf L^{\top}\blf D\blf M,\mathbf{I}_N) +d_{N+1}(\blf t_L\bsym{\mu}^{\o\top}\overline{\blf M}_t+\blf T_{LM})
\end{split}
\end{align}
where $\blf M=[\blf I,\ldots,\blf A^M]$, $\blf M_t=[\blf t,\blf A\blf t,\ldots,\blf A^M\blf t]$, $\overline{\blf M}_t=[\blf 0, \blf t,\blf A\blf t,\ldots,\blf A^{M-1}\blf t]$, $\blf t_L=[t_+,\blf 0_L]$, and $\blf T_{LM}\in\mathbb{R}^{L+1\times M+1}$ with  $(t_+^2+\sigma^2)$ in location $(1,1)$ and zero elsewhere
\begin{align}\label{eq.Delta22}
    \begin{split}
      \boldsymbol{\Delta}_{22}=&\blf M^{\top}_x\blf D\blf M_x+\sigma^2\blktr(\blf M^{\top}\blf{D}_N\blf M,\mathbf{I}_N)\\
    &+d_{N+1}\big(\blktr(\overline{\blf M}^{\top}\blf R^\o\overline{\blf M},(\sigma^2\mathbf{I}+\mathbf{t}\mathbf{t}^{\top}))\\
    &+\overline{\blf M}_t^{\top}\bsym{\mu}^{\o}\blf{t}_M^{\top}+\blf{t}_M\bsym{\mu}^{\o\top}\overline{\blf M}_t+\textnormal{diag}(t_+^2+\sigma^2,\blf{0}_M)\big)
    \end{split}
\end{align}
where $\overline{\blf M}=[\blf 0, \blf I,\ldots,\blf A^{M-1}]$, $\blf R^\o=\bsym{\Sigma}^\o+\bsym{\mu}^\o\bsym{\mu}^{\o\top}$, and $\blf t_M=[t_+,\blf 0_M]$. Vector $\boldsymbol{\theta}\in \mathbb{R}^{L+M+2}$ is of the form
\begin{align}\label{eq.theta}
    \begin{split}
    \boldsymbol{\theta}=\begin{bmatrix}
    (\blf L_t+t_+\overline{\blf L}_{\mu^\i})^{\top}\blf t+t_+\blf t_L\\
    \blf M_t^{\top}\mathbf{t}+ t_+\overline{\blf M}_t^{\top}\bsym{\mu}^\i+t_+\blf t_M
    \end{bmatrix}.
    \end{split}
\end{align}
\end{proposition}

\textit{Proof.} See the appendix in the Supplementary Material.\hfill\qed

{The MSE in \eqref{sure} is governed by the interactions between the statistics of the attachment vectors, and those of the percolated signals $\blf t$ and $\bsym{\mu}^\i$.
} Using it as the cost, problem \eqref{pbasic} becomes
\begin{align}\label{pde-noising}
\begin{split}
\underset{\mathbf{h} = [\mathbf{h}^{\i, \top},\!\! \mathbf{h}^{\o, \top}]^\top}{\text{min }} \frac{1}{2\gamma} \textnormal{MSE}_{\mathbf{D}}(\mathbf{h})\!+\!\frac{1}{2\alpha}||\mathbf{h}^\i||_2^2\!+\!\frac{1}{2(1-\alpha)}||\mathbf{h}^\o||_2^2.\\
\end{split}
\end{align}
Scalar $\gamma \!>\!0$ controls how much we want to reduce the MSE over the nodes in $\mathbf{D}$; for $\gamma \to 0$ the importance of minimizing the MSE increases, while for $\gamma \to \infty$ it decreases. Instead, scalar $\alpha \in ]0,1[$ controls the role of the filters $\mathbf{H}^\i(\!\mathbf{A}^{\i}_{+}\!)$ and $\mathbf{H}^\o_{+}(\!\mathbf{A}^\o_{+}\!)$ in \eqref{ncr}. For $\alpha \to 0$ we prioritise more the filter over graph $\mathcal{G}^{\i}$; i.e., leverage the information on the existing nodes $\cal V$ towards the incoming node $v_+$. And for $\alpha \to 1$ we prioritise the filter over graph $\mathcal{G}^{\o}$; i.e.,  leverage the information on the incoming node $v_+$ towards the existing nodes $\cal V$. Note that such a formulation would work for any additive noise or attachment model as long as respective parameters are known.
\par Problem \eqref{pde-noising} is quadratic and convex only if matrix $\boldsymbol{\Delta}$ is positive semi-definite (PSD). However, proving the latter is challenging because of the structure of this matrix; hence, we can find local minima via descent algorithms \cite{boyd2004convex}. But since we estimate $\boldsymbol{\Delta}$ from the training set $\mathcal{T}$, we can check if it is PSD and for a positive outcome we can find the closed-form solution for \eqref{pde-noising}
\begin{align}\label{pete}
    \begin{split}
        {\blf h}^{\star}=\big(\bsym{\Delta}_{\cal T}+2\gamma\boldsymbol{\Lambda}\big)^{-1}\bsym{\theta}_{\cal T}
    \end{split}
\end{align}
where matrix $\boldsymbol{\Lambda}\!\!\!\!\!=\![
1/2\alpha\mathbf{I}_{L+1},\mathbf{0};
\mathbf{0},1/2(1-\alpha)\mathbf{I}_{M+1}
]\!\in\!\bb R^{(M+L+2)\times(M+L+2)}$ and subscript $_{\mathcal{T}}$ indicates that these quantities are estimated from data. 

\vspace{-5mm}
\subsection{Semi-supervised Learning}\label{subsec_SSL}
As second task, we perform inductive semi-supervised learning (SSL) \cite{zhu2005semi}. As in \cite{sandryhaila_discrete_2013,berberidis2018adaptive,chen_semi-supervised_2014} we use graph filters for such a task but now operating over the expanded graph. Specifically, we consider a binary classification problem with sparse label target vector $\mathbf{t}_+$ such that $[\mathbf{t}_+]_n = \pm1$ if node $n \in \mathcal{V} \cup v_+$ is labelled, or zero if unlabeled. As learning cost for problem \eqref{pbasic}, we consider the label fitting $ \textnormal{MSE}_{\mathbf{D}}(\mathbf{h}) = {\bb E}[||\mathbf{y}_+-\mathbf{t}_+||_\mathbf{D}^2]={\bb E}[(\mathbf{y}_+-\mathbf{t}_+)^{\top}\mathbf{D}(\mathbf{y}_+-\mathbf{t}_+)]$ which is a typical convex approach for graph-based SSL with satisfactory results \cite{berberidis2018adaptive}. Then, in addition to regularizing the problem w.r.t. the $\ell_2-$norm of the filter coefficients, we consider also graph-regularizers via the expected discrete $2-$Dirichlet form w.r.t. both graphs $\cal G_+^\i$ and $\cal G_+^\o$; i.e., $\bb E[{S}_2(\blf y^\i_+)] = {\bb E}[||{\bf y}^\i_+ - {\bf A}^\i_+{\bf y}^\i_+||_2^2]$ and $\bb E[{S}_2(\blf y^\o_+)] = {\bb E}[||{\bf y}^\o_+ - {\bf A}^\o_+{\bf y}^\o_+||_2^2]$, respectively. The latter ensures that each filter output is a smooth over directed graphs, which has been validated for graph-based SSL \cite{sandryhaila_discrete_2013,berberidis2018adaptive,chen_semi-supervised_2014,zhu2005semi,smola2003kernels,zhou2004regularization}. Then, Problem \eqref{pbasic} becomes
\begin{align}\label{pssl}
    \begin{split}
    &\underset{\mathbf{h} = [\mathbf{h}^{\i, \top}, \mathbf{h}^{\o, \top}]^\top}{\text{min }} \frac{1}{2\gamma}\textnormal{MSE}_{\mathbf{D}}(\mathbf{h})+\frac{1}{2\alpha}||\mathbf{h}^\i||_2^2+\frac{1}{2(1-\alpha)}||\mathbf{h}^o||_2^2\\
    &\quad+\frac{1}{2\beta}\bb E[||{\bf y}^\i_+ - {\bf A}^\i_+{\bf y}^\i_+||_2^2]+\frac{1}{2(1-\beta)}\bb E[||{\bf y}^\o_+ - {\bf A}^\o_+{\bf y}^\o_+||_2^2]
    \end{split}
    \end{align}
where again $\gamma > 0$ controls the trade-off between the fitting term and the regularizer,  $\alpha \in ]0,1[$ controls the roles of the filter behavior over graphs $\cal G_+^\i$ and $\cal G_+^\o$ [cf. \eqref{pde-noising}], and $\beta \in ]0,1[$ controls now the filter output smoothness w.r.t. graphs $\cal G_+^\i$ and $\cal G_+^\o$. For $\beta \to 0$, we bias filter $\mathbf{H}^\i(\mathbf{A}^{\i}_{+})$ to give an output ${\bf y}^\i_+$ that is smooth over graph $\cal G_+^\i$ and to \emph{ignore} the behavior of filter output ${\bf y}^\o_+$ over graph $\cal G_+^\o$. This may be useful when the connectivity model of $v_+$ respects the clustering structure of $\cal G$.
The opposite trend is observed for $\beta \to 1$.

The MSE in \eqref{pssl} is of the form \eqref{sure} and encompasses both SSL cases with clean labels ($\sigma^2 = 0$) and noisy labels ($\sigma^2 > 0$). In \eqref{pssl} also have the expected signal $2-$Dirichlet form that influences the filter behavior. The following proposition quantifies it.

\begin{proposition}\label{Prop 2}
    \textit{Given the setting of Proposition \ref{zion} and considering that $\blf x_+=\blf t_+$, the $2-$Dirichlet forms of the filter outputs $\blf y_+^\i$ and $\blf y_+^\o$ over graphs $\cal G^\i_+$ and $\cal G^\i_+$ are respectively}
    \begin{equation}\label{eq.quadForms}
       \bb E[{S}_2(\blf y^\i_+)]=\blf h^{\i\top}\bsym{\Psi^\i} \blf h^{\i}~~~\textnormal{and}~~~\bb E[{S}_2(\blf y^\o_+)]=\blf h^{\o\top}\bsym{\Psi^\o} \blf h^{\o}
    \end{equation}
    where
    \begin{align}\label{eq.psii}
    \begin{split}
    \bsym{\Psi^\i}&=(\blf L_t+t_+\overline{\blf L}_{\mu^\i})^{\top}\boldsymbol{\Gamma}(\blf L_t\!+\!t_+\overline{\blf L}_{\mu^\i})+t_+^2\blktr(\overline{\blf L}^{\top}\boldsymbol{\Gamma}\overline{\blf L},\bsym{\Sigma}^\i)\\
    &-2\blf t_L\big(\bsym{\mu}^{\i\top}\blf L_t\!+\!t_+\bsym{\mu}^{\i\top}\overline{\blf L}_{\mu^\i}+t_+\blktr(\overline{\blf L},\bsym{\Sigma}^\i)\big)\\
&-\!\blf t_L(\bsym{\mu}^{\i\top}\blf A\blf L_t\!-\!t_+\bsym{\mu}^{\i\top}\blf A\overline{\blf L}_{\mu^\i}\!-\!t_+\blktr(\blf A\overline{\blf L},\bsym{\Sigma}^\i))\\
&+(\blf w^{\top}\blf p^\i+1)\textnormal{diag}(t_+,\blf 0)
\end{split}
\end{align}\vskip -0.25cm
\begin{align}\label{eq.psio}
\begin{split}
    &\bsym{\Psi^\o}=\blf M_t^{\top}(\boldsymbol{\Gamma}+\blf R^\o)\blf M_t-\blf M_t^{\top}\blf R^\o\overline{\blf M}_t+\blf M_t^{\top}\bsym{\mu}^\o\blf t_M^{\top}-\overline{\blf M}_t^{\top}\blf R^\o\blf M_t\\
    &+\blf t_M\bsym{\mu}^{\o\top}\blf M_t+\overline{\blf M}_t^{\top}\blf R^\o\overline{\blf M}_t+\overline{\blf M}_t^{\top}\bsym{\mu}^{\o}\blf t_M^{\top}+\blf t_M\bsym{\mu}^{\o\top}\overline{\blf M}_t
  +\blf t_M\blf t_M^{\top}
  \end{split}
\end{align}

\noindent are matrices that capture the attachment patterns and label propagation on the edges of the incoming node and $\boldsymbol{\Gamma} = (\blf I - \blf A)^{\top}(\blf I - \blf A)$.

\end{proposition}

\textit{Proof.} See the appendix in the Supplementary Material.\hfill\qed

The expected $2-$Dirichlet forms depends on the attachment statistics in two ways: first, the expected attachments $\boldsymbol{\mu}^{\i}$ and $\boldsymbol{\mu}^{\o}$ control the label percolation from and towards the incoming node; second the in-attachment covariance $\bsym{\Sigma}^\i$ and the out-attachment covariance $\bsym{\Sigma}^\o$ influence the percolated labels through $\overline{\blf L}$ and $\blf A$ and $\blf M_t$ and $\overline{\blf M}_t$. Using then \eqref{eq.quadForms} in \eqref{pssl}, we get
\begin{align}\label{psslnew}
\begin{split}
    \underset{\mathbf{h} = [\mathbf{h}^{\i, \top}, \mathbf{h}^{\o, \top}]^\top}{\text{min }} &\frac{1}{2\gamma}\textnormal{MSE}_{\mathbf{D}}(\mathbf{h})+\blf h^{\top}\bsym{\Lambda}\blf h+\blf h^{\top}\bsym{\Omega}\blf h
\end{split}
\end{align}
where $\bsym{\Lambda}$ is defined in \eqref{pete} and
$\bsym{\Omega}= [1/2\beta\bsym{\Psi}^\i,\blf{0}; \blf{0},1/2(1-\beta)\bsym{\Psi}^\o]$ is an $(M\!+\!L\!+\!2)\!\times\!(M\!+\!L\!+\!2)$ matrix.
%
As for \eqref{pssl} proving convexity for \eqref{psslnew} is challenging but solvable with descent algorithms. And if empirically we observe that the matrix in the quadratic form of $\blf h$, $\boldsymbol{\Delta}_{\cal T}+\bsym{\Lambda}+\bsym{\Omega}$ is PSD, the solution of \eqref{psslnew} is given by
\begin{align}
    \begin{split}
        {\blf h}^{\star}=\big({\boldsymbol{\Delta}_{\cal T}}+2\gamma(\boldsymbol{\Lambda}+\bsym{\Omega}_{\cal T}+\bsym{\Omega}_{\cal T}^{\top})\big)^{-1}\boldsymbol{\theta}_{\cal T}
    \end{split}
\end{align}
where again the subscript $\mathcal{T}$ indicates that the respective quantities are estimated from data. 


\definecolor{Gray}{gray}{0.9}
\begin{table*}[t]
\centering
\caption{\label{MSE comparison}NMSE over all nodes and $\text{NMSE}_+$ of the different models for different SNRs.}
\begin{tabular}{c c c c c c c c c c c c c}
\hline\hline
& \multicolumn{6}{c}{\footnotesize{Barabassi-Albert}} & \multicolumn{6}{c}{\footnotesize{NOAA}}\\ \cmidrule(l{20pt}r{25pt}){2-7} \cmidrule(l{20pt}r{29pt}){8-13}
 & \multicolumn{2}{c}{\footnotesize{SNR 5dB}} & \multicolumn{2}{c}{\footnotesize{SNR 10dB}} & \multicolumn{2}{c}{\footnotesize{SNR 20dB}}& \multicolumn{2}{c}{\footnotesize{SNR 5dB}} & \multicolumn{2}{c}{\footnotesize{SNR 10dB}} & \multicolumn{2}{c}{\footnotesize{SNR 20dB}}\\ \cmidrule(l{5pt}r{10pt}){2-7}\cmidrule(l{50pt}r{10pt}){7-13}
\footnotesize{Rule} & \footnotesize{NMSE} & $\text{\footnotesize{NMSE}}_+$ & \footnotesize{NMSE} & $\text{\footnotesize{NMSE}}_+$ & \footnotesize{NMSE} & $\text{\footnotesize{NMSE}}_+$ & \footnotesize{NMSE} & $\text{\footnotesize{NMSE}}_+$ & \footnotesize{NMSE} & $\text{\footnotesize{NMSE}}_+$ & \footnotesize{NMSE} & $\text{\footnotesize{NMSE}}_+$ \\\hline
\rowcolor{Gray}
\footnotesize{\textbf{Prop.}} & $8\times 10^{-2}$ & 0.79 & 0.073 & 0.8 & $7\times10^{-4}$ & 0.54 & 0.103  & 0.156  & 0.054  & 0.110  & $9\times 10^{-3}$  &  $1.3\times 10^{-2}$ \\
\footnotesize{\textbf{KC$_1$}} & $7\times 10^{-2}$ & 0.26 & 0.069 & 0.16 & $4\times10^{-4}$ & $9\times10^{-4}$ & 0.063 & 0.136 & 0.03 & 0.073 & $6\times 10^{-3}$ & $1.2\times 10^{-2}$ \\
\rowcolor{Gray}
\footnotesize{\textbf{KC$_2$}} & $8\times 10^{-2}$ & 0.46 & 0.07 & 0.40  &$5\times10^{-4}$ & 0.03 & 0.103 & 0.09 & 0.054 & 0.07  & $9\times 10^{-3}$ & $1.2\times 10^{-2}$ \\
\footnotesize{\textbf{IT}} & $7\times 10^{-2}$ & 3.61 & 0.069 & 3.5 & $7\times10^{-4}$ & 3.7 & 0.076 & 0.204 & 0.041 & 0.122 & $8\times 10^{-3}$ & $1.4\times 10^{-2}$ \\
\hline\hline
\end{tabular}\vskip-.5cm
\label{synthetic de-noising}
\end{table*}

\vspace{-5mm}
\section{Numerical Results}\label{Section Numerical}

This section compares the proposed method with competing alternatives to illustrate the trade-offs inherent to graph filtering over expanding graphs with synthetic and real data. Our numerical tests have been focused to answer the following research questions:
\begin{enumerate}[label=\textbf{RQ.\arabic*.}, start = 1]
\item How does the proposed approach compare with baselines that utilize the known attachment?
\end{enumerate}
That is, we want to understand to what extent the proposed empirical learning framework compensates for the ignorance of the true connection. To answer this question, we compare with two baselines:
\begin{enumerate}
\item \textit{Single filter with known connectivity} (\textbf{KC$_1$}): This is the intuitive solution where the incoming node $v_+$ connects to the nodes in $\mathcal{V}$ forming a single graph $\mathcal{G}_+ = (\mathcal{V} \cup v_+, \mathcal{E}_+)$, in which set $\mathcal{E}_+$ collects both the \emph{known} incoming and outgoing edges w.r.t. $v_+$. Then, a single filter is trained on $\mathcal{G}_+$ as conventionally done by the state-of-the-art. This comparison {validates} the proposed scheme over the conventional strategy.
\item \textit{Filter bank with known connectivity} (\textbf{KC$_2$}): This is the proposed filter bank scheme in \eqref{ncr} with the \emph{known} connectivity of node $v_+$. The rationale behind this choice is to factorize the filter degrees of freedom since $\textbf{KC$_1$}$ employs a single filter and to highlight better the role of the topology.
\end{enumerate}
\begin{enumerate}[label=\textbf{RQ.\arabic*.}, start = 2]
\item How much does the information of the attached node contribute to the task performance over the existing graph?
\end{enumerate}
We want to understand if the proposed approach exploits the incoming node signal without knowing the topology to improve the task over the existing graph instead of ignoring such information.

\begin{enumerate}[label=\textbf{RQ.\arabic*.}, start = 3]
\item How does the proposed model compare on the incoming node w.r.t. inductive graph filtering?
\end{enumerate}
Since graph filters have inductive bias capabilities \cite{gama2020graphs}, they can be learned on the exiting graph $\mathcal{G}$ and then transferred to expanded graphs without retraining. We want to understand if learning with a stochastic model is more beneficial than transference. To answer RQ.2 and RQ.3, we compare with:
\begin{enumerate}[label=\arabic*., start = 3]
\item \textit{Inductive transference} (\textbf{IT}): I.e., we employ a single filter to solve the task over the existing graph and transfer it on the expanded graph under the same attachment model.
\end{enumerate}


For all experiments, the incoming node attaches to the existing nodes ($\cal G_+^{\i}$) uniformly at random with $\blf p^{\i}=\mathbf{1}_N/N$, and have edges landing at itself ($\cal G_+^{\o}$) with a preferential attachment $\bf p^{\o}=\mathbf{d}/\mathbf{1}^{\top}\mathbf{d}$ where $\blf d$ is the degree vector; i.e., they are likelier to form links with nodes having a higher degree. These standard attachment rules have been observed in the  study of evolving real world networks. The covariance matrices are estimated from $10,000$ generated samples of their respective attachment vectors. For simplicity, we set the expanded graph weights $\mathbf{w}^{\i} \!=\! \mathbf{w}^{\o} \!=\! w\mathbf{1}$ with $w$ being the median of the non-zero existing edge weights in $\cal G$. We fixed the filter orders to $L\!=\!M\!=\!4$ and considered also a filter order of four for \textbf{KC}$_1$ and \textbf{IT}. We performed a $70-30$ train-test data split and selected parameters $\gamma \in [10^{-3},10]$, $\alpha, \beta \in ]0, 1[$ via five-fold cross-validation. We averaged the testing performance over $100$ realizations per test node.


\vspace{-5mm}
\subsection{Denoising}

Following the paper outline, we first answer the RQs for the de-noising task over a Barabasi-Albert (BA) graph model and the NOAA temperature data-set \cite{arguez2012noaa}.

\smallskip
\noindent\textbf{Experimental setup.} For the BA model, we considered an existing graph of $100$ nodes and $1000$ incoming node realizations. For each realization, we generated a bandlimited graph signal by randomly mixing the first ten eigenvectors with the smallest variation of $\mathbf{A}_+ \in \mathbb{R}^{101\times 101}$ \cite{sandryhaila_discrete_2013}. For the NOAA data set, we considered hourly temperature recordings over $109$ stations across the continental U.S. in $2010$. We built a five nearest neighbors (5NN) graph $\mathcal{G}$ of $N = 100$ random stations as in \cite{mei2016signal}\cite{isufi2019forecasting}. We treated the remaining nodes as incoming, each forming 5NN on $\cal G^{\i}$ and 5NN on  $\cal G^{\o}$. We considered $200$ hours, yielding $1800$ incoming data samples.

We corrupted the true signals with Gaussian noise of SNRs $\in \{5\textnormal{dB}, 10\textnormal{dB}, 20\textnormal{dB}\}$. We measured the  recovery performance over all existing and incoming nodes through the normalized mean squared error $\text{NMSE}={||\mathbf{y}_+-\mathbf{t}_+||_2^2}/{||\mathbf{t}_+||_2^2}$ and we also measured the NMSE only at the incoming node and denote it as $\text{NMSE}_+$.

\smallskip
\noindent\textbf{Observations.} Table~\ref{synthetic de-noising} reports the denoising performance on both datasets. Overall, we observe that the proposed approach compares well with the two baselines relying on the exact topology (\textbf{KC$_1$} and \textbf{KC$_2$}). As regards the performance at the incoming node $\text{NMSE}_+$, we see that not knowing the topology leads to a worse performance. However, we see that in the NOAA dataset the gap is much smaller; a potential explanation for this is may be in the NN nature of the graph. Regarding then the last two research questions, we see that the proposed approach performs comparably well w.r.t. \textbf{IT} on the existing graphs but outperforms it by a margin when it comes to the performance of the incoming node ($\text{NMSE}_+$). In turn, such findings show the advantages of the proposed scheme to keep a comparable performance with baselines relying on the exact topology and to improve substantially w.r.t. methods relying only on transference.

\vspace{-5mm}
\subsection{Semi-supervised Learning}

For SSL, we consider a synthetic sensor network graph from the GSP toolbox \cite{perraudin2014gspbox} and the political blog network \cite{adamic2005political}.

\smallskip
\noindent\textbf{Experimental setup.} For the sensor network, the existing graph $\mathcal{G}$ has $N=200$ nodes that are clustered into two classes ($\pm1$) via spectral clustering to create the ground-truth. The training set $\mathcal{T}$ comprises $500$ realizations of incoming nodes each making the same number of incoming and outgoing as the median degree of $\mathcal{G}$. The ground-truth label at the incoming node is assigned based on the class that has more edges with $v_+$. For the blog network, we considered $1222$ blogs as nodes of a graph with directed edges being the hyperlinks between blogs and labels being their political orientation ($+1$ conservative vs. $-1$ liberal). We built a connected existing graph $\mathcal{G}$ of $N = 622$ blogs with a balanced number of nodes per class. The remaining $600$ blogs are treated as incoming nodes.

In both settings, we use only $10\%$ of the labels in $\mathcal{G}$ and aim at inferring the missing labels in this graph by using also the information from the incoming node. These labels act also as the graph signal $[\mathbf{x}_+]_n=\pm1$ for a labelled node and $[\mathbf{x}_+]_n=0$ if unlabeled. We also considered two settings: first, all incoming nodes in the training set have labels (fully labelled), which allows identifying if the additional label contributes to the SSL task on $\mathcal{G}$; second, only half of the incoming nodes have labels ($50\%$ labelled), which adheres more to a real scenario where some of the incoming nodes are unlabeled. For the $\textbf{IT}$ baseline, we solve the corresponding filters using \cite{sandryhaila_discrete_2013}, while for $\textbf{KC}_1$ and $\textbf{KC}_2$ we use the true connections. During training, standard SSL requires evaluating the loss at the nodes with available labels. Hence, when an incoming node has no label, we cannot account for its importance during training. Consequently, SSL models cannot predict labels when we do not know the connectivity. Thus, we measure only the performance of the existing nodes.
\begin{table}[t!]
\centering
\caption{\label{MSE comparison} Average ($\pm$ std.) SSL error for the sensor network.}
\begin{tabular}{c c c} 
\hline\hline
Error (\%) & Fully labelled & 50 \% labelled \\\hline
\rowcolor{Gray}
\textbf{Prop.} & 4.64~($\pm3.43$) & 4.8~($\pm2.59$)\\
\textbf{KC$_1$} & 5.35~($\pm 3.36$) & 5.6~($\pm2.6$)\\
\rowcolor{Gray}
\textbf{KC$_2$} & 4.62~($\pm3.41$) & 4.9~($\pm2.62$)\\
\textbf{IT} & 6.12~($\pm3.78$) & 5.3~($\pm2.63$)\\
\hline\hline
\end{tabular}\vskip-.5cm
\label{ssl synthetic}
\end{table}

\begin{table}[t!]
\centering
\caption{\label{MSE comparison} Average ($\pm$ std.) SSL  error for the blog network.}
\begin{tabular}{c c c} 
\hline\hline
Error (\%) & Fully labelled & 50 \% labelled \\\hline
\rowcolor{Gray}
\textbf{Prop.} & 2.8 ~($\pm0.4$) & 2.56~($\pm0.75$)\\
\textbf{KC$_1$} & 2.82 ~($\pm0.83$) & 2.42~($\pm0.59$)\\
\rowcolor{Gray}
\textbf{KC$_2$} &  2.8 ~($\pm0.4$) & 2.56~($\pm0.75$)\\
\textbf{IT} & 12.2 ~($\pm18$) & 6.58~($\pm0.59$)\\
\hline\hline
\end{tabular}\vskip-.5cm
\label{ssl real}
\end{table}

\smallskip
\noindent\textbf{Observations.} Tables~\ref{ssl synthetic} and \ref{ssl real} report the classification errors for the sensor and blog networks, respectively. The proposed approach achieves a comparable statistical performance with the two baselines (\textbf{KC$_1$} and \textbf{KC$_2$}) that rely on the exact topology. This suggests that controlling the information in-flow and out-flow with a filter bank compensates effectively for the exact topology ignorance. The proposed approach reduces the error substantially compared to $\textbf{IT}$.

From these results, we also observe the models tend to perform better when $50\%$ of the labels are present. We have identified two factors for this. First, some of the incoming nodes form misleading connections with both clusters. Hence, when their label diffuses it hampers the classification performance on the opposite cluster. Instead, when these nodes have no label they do influence the opposite class. This trend is observed also for $\textbf{KC}_1$ and $\textbf{KC}_2$, which shows that these wrong connections are present in the dataset. In the blog network, these are blogs with an unclear political position and have linked both with liberals and conservative groups \cite{adamic2005political}. Instead, in the sensor network, we do not see such a trend because nodes are better clustered. Second, this two-class classification problem has labels $\pm1$ and we use the MSE as a criterion. Hence, the term $t_+^2=1$ affects the costs when the incoming node is present $[\text{cf. Prop.} \ref{zion},\ref{Prop 2}]$ and does not help discriminating irrespective of the class. Thus, we conclude that when dealing with SSL classification in expanding graphs, the connectivity model plays also a central role in the performance.

\smallskip
\vspace{-5mm}
\section{Conclusion}\label{Section Conclusion}\vskip -1mm
We proposed a method to filter signals over expanding graphs by relying only on their attachment model connectivity. We used a stochastic model where incoming nodes connect to the existing graph, forming two directed graphs. A pair of graph filters, one for each graph, then process the expanded graph signal. To learn the filter parameters, we performed empirical risk minimisation for graph signal de-noising and graph semi-supervised learning. Numerical results over synthetic and real data show the proposed approach compares well with baselines relying on exact topology and outperforms the current solution relying on filter transference. However, the performance is strongly dependent on a fixed attachment model, prone to model mismatch. Hence, potential future works may consider a joint filter and graph learning framework for expanding graphs.

\bibliographystyle{IEEE.bst}
\bibliography{refs}
\clearpage
\section*{Supplementary Material}\label{appendix}
\renewcommand{\theequation}{S\arabic{equation}}
\setcounter{equation}{0}

This document contains the proofs  of the main claims of the paper \textit{Graph filtering over expanding graphs}. 

\subsection*{Proof of Lemma \ref{velvet}}\label{honesthearts}
Substituting $\blf{L}_x=\blf{L}(\mathbf{I}_{L+1}\otimes\mathbf{x})$ and $\blf{M}_x=\blf{M}(\mathbf{I}_{M+1}\otimes\mathbf{x})$ into $\bb E[\blf L_x^{\top}\blf{C}\blf M_x]$, we get
\begin{equation}\label{deathclaw}
   \bb E[\blf L_x^{\top}\blf{C}\blf M_x]=\bb E[(\mathbf{I}_{L+1}\otimes\mathbf{x})^{\top}\blf L^{\top}\blf{C}\blf M(\mathbf{I}_{M+1}\otimes\mathbf{x})].
\end{equation}
Substituting further $\blf L=[\blf I, \blf A,\ldots, \blf A^L]$, $\blf M=[\blf I, \blf A,\ldots, \blf A^M]$, the $(i,j)$th block of $\blf L^{\top}\blf C \blf M$, is
\begin{equation}\label{eq.dummy0L1}
    [\blf L^{\top}\blf C \blf M]_{ij} \!=\! \blf{A}^{i\!-\!1}\blf{C}\blf{A}^{j\!-\!1}~\textnormal{for}~ \{i,j\} = 1, \ldots, \{L\!+\!1, M\!+\!1\}.
\end{equation}
%
%
Incorporating the Kronecker products involving $\blf x$, we further write the $(i,j)$th entry of \eqref{deathclaw} as
\begin{equation}\label{eq.dummy1L1}
    \bb E[\blf L_x^{\top}\blf{C}\blf M_x]_{ij} = \bb E [\blf x^{\top}\blf A^{i-1}\blf C\blf A^{j-1} \blf x]
\end{equation}
since the expectation acts element-wise.
%
%
Since the expectation argument is a scalar, we bring in the trace operator and leverage its cyclic property $\tr({\blf X \blf Y \blf Z})=\tr({\blf Z \blf X \blf Y})$ to write
\begin{align}\label{123}
    \begin{split}
       \bb E[\blf x^{\top}\blf A^{i-1}\blf C\blf A^{j-1} \blf x]= \bb E[\tr(\blf x \blf x^{\top}\blf A^{i-1}\blf{C}\blf A^{j-1})]
    \end{split}
\end{align}
where remark the only random variable in \eqref{123} is $\blf{x}$.
%
Substituting then $\bb E[\blf x \blf x^{\top}]=\blf t \blf t^{\top}+\sigma^2\blf I_N$ in \eqref{123}, we get
\begin{align}\label{barkscorpion}
    \begin{split}
    \bb E[\blf L_x^{\top}\blf{C}\blf M_x]_{ij} =\tr(\blf t \blf t^{\top}\!\!\blf A^{i\!-\!1}\blf{C}\blf A^{j\!-\!1})
    +\sigma^2\tr(\blf A^{i-1}\blf{C}\blf A^{j-1})
    \end{split}
\end{align}
The first term in the R.H.S. of \eqref{barkscorpion} is $[\blf L_t^{\top}\blf{C}\blf M_t]_{ij} = \blf t^{\top}\blf A^{i-1}\blf C\blf A^{j-1} \blf t$ with $\blf L_t=\blf L_x|_{x=t}$ and $\blf M_t=\blf M_x|_{x=t}$. Instead for the second term $\sigma^2\tr(\blf A^{i-1}\blf{C}\blf A^{j-1})$, we leverage \eqref{eq.dummy0L1} and Def. \ref{hmm} and note that it  is the $(i,j)$th element of $\blktr(\blf L^{\top}\blf{C}\blf M,\sigma^2\blf{I}_N)$.
%
%
Thus, the $(i,j)$th element of $\bb E [\blf L_x^{\top}\blf{C}\blf M_x]$ is the sum of the $(i,j)$th element of these two matrices, proving the Lemma.\hfill\qed

\subsection*{Proof of Proposition \ref{zion}}\label{saltuponwounds}
Expanding the MSE definition we get
\begin{align}\label{227}
    \begin{split}
      \mathbb{E}[||\mathbf{W}{_+}\blf h-\mathbf{t}_{\!+}||_{\blf D_+}^2]\!=&\!\blf h^{\!\top}\bb E[\blf{W}_+^{\!\top}\blf{D_+} \blf{W}_+]\blf{h} \!+2\blf h^{\!\top}\bb{E}[\bf W_+^{\!\top}\blf D_+\bf t_+]\!\\
      &+\!\bf t_+^{\!\top}\bf D_+\bf t_+ 
    \end{split}
\end{align}
The first term contains the matrix $\boldsymbol{\Delta} := \bb E[\bf W_+^{\top}\bf{D}\bf W_+]$. Substituting $\blf W_+$ [cf. \eqref{eq.filt_compact}], we the expectation argument becomes
%
%
\begin{align}\label{rad}
\begin{split}
&\blf W_+^{\top}\blf D_+\blf W_+= \begin{bmatrix}
        \widehat{\mathbf{L}}_x^{\top} &\mathbf{x}_L \\
        \mathbf{M}_x^{\top} & \widehat{\mathbf{m}}_{x}\\
\end{bmatrix}\begin{bmatrix}
        \blf D &\blf 0\\
       \blf 0 & d_{N+1}\\
\end{bmatrix}\begin{bmatrix}
        \widehat{\mathbf{L}}_x &\mathbf{M}_x\\
        \mathbf{x}_L^{\top} & \widehat{\mathbf{m}}_{x}^{\top}\\
\end{bmatrix}\\
&=\begin{bmatrix}
\widehat{\mathbf{L}}_x^{\top}\blf D\widehat{\mathbf{L}}_x+d_{N+1}\mathbf{x}_L \mathbf{x}_L^{\top} & \widehat{\mathbf{L}}_x^{\top}\blf D\mathbf{M}_x+d_{N+1}\mathbf{x}_L\widehat{\mathbf{m}}_{x}^{\top}\\
\mathbf{M}_x^{\top}\blf D\widehat{\mathbf{L}}_x+d_{N+1}\widehat{\mathbf{m}}_{x}\mathbf{x}_L^{\top} & \mathbf{M}_x^{\top}\blf D\mathbf{M}_x+d_{N+1}\widehat{\mathbf{m}}_{x}\widehat{\mathbf{m}}_{x}^{\top}
\end{bmatrix}
\end{split}
\end{align}
which are related to the four blocks appearing in $\boldsymbol{\Delta}$ and where $\blf D=\textnormal{diag}(d_1,\ldots,d_N)$.

\smallskip
\noindent\textbf{$\bsym{\Delta}_{11}$.} The first block matrix in \eqref{rad} is $\bsym{\Delta}_{11}:= \bb E [\widehat{\mathbf{L}}_x^{\top}\blf D\widehat{\mathbf{L}}_x+d_{N+1}\mathbf{x}_L \mathbf{x}_L^{\top}].$
Substituting $\widehat{\mathbf{L}}_x=\mathbf{L}_x+x_+\overline{\mathbf{L}}_b$, we get
\begin{align}
   \bb E [\widehat{\mathbf{L}}_x^{\top}\blf D\widehat{\mathbf{L}}_x]=\bb E [(\mathbf{L}_x+x_+\overline{\mathbf{L}}_b)^{\top}\blf D(\mathbf{L}_x+x_+\overline{\mathbf{L}}_b)]
\end{align}
which is further composed of the following four terms:
\begin{itemize}
    \item $\bb E[\mathbf{L}_x^{\top}\blf D\mathbf{L}_x]=\mathbf{L}_t^{\top}\blf D\mathbf{L}_t+\sigma^2\blktr(\blf L^{\top}\blf D\blf L,\blf I_{N})$, which follows directly from Lemma \ref{velvet};
\item $\bb E [x_+\mathbf{L}_x^{\top}\blf D\overline{\mathbf{L}}_b] = t_+\mathbf{L}_t^{\top}\blf D\overline{\mathbf{L}}_{\mu^\i}$ given the noise and the attachments are independent of each other;
\item $\bb E [x_+\overline{\mathbf{L}}_b^{\top}\blf D\blf L_x] = t_+\overline{\mathbf{L}}_{\mu^\i}^{\top}\blf D\blf L_t$ under the same independence considerations;
\item $\bb E [x_+^2\overline{\mathbf{L}}_b^{\top}\blf D\overline{\mathbf{L}}_b]$. Under the independence between $x_+$ and $\blf b_+^\i$, and by using Lemma \ref{velvet} on $\overline{\mathbf{L}}_b^{\top}\blf D\overline{\mathbf{L}}_b$, we get
\begin{equation}
    \bb E [x_+^2\overline{\mathbf{L}}_b^{\!\top}\blf D\overline{\mathbf{L}}_b] \!=\! (t_+^2+\sigma^2)\big(\overline{\blf L}^{\!\top}_{\mu^\i}\blf D\overline{\blf L}_{\mu^\i}+\blktr(\overline{\blf L}^{\!\top}\blf D\overline{\blf L},\boldsymbol{\Sigma}^{\i})\big)
\end{equation}
where we also used the identity $\bb E[\blf b_+^\i \blf b_+^{\i\top}]=\bsym{\mu}^\i\bsym{\mu}^{\i\top}+\bsym{\Sigma}^\i$.
\end{itemize}
%
%
%
%
In the expression of $\bsym{\Delta}_{11}$ we also have the term $\bb E [d_{N+1}\mathbf{x}_L\mathbf{x}_L^{\top}] = d_{N+1}\textnormal{diag}(t_+^2+\sigma^2,\blf{0}_L)$, which holds because $\mathbf{x}_L = [x_+,\blf 0_L]$. Combining these, we get expression \eqref{eq.Delta11} for $\bsym{\Delta}_{11}$.

\smallskip
\noindent\textbf{$\bsym{\Delta}_{12}$.} The second block matrix in \eqref{rad} is $\bsym{\Delta}_{12} := \bb E [\widehat{\mathbf{L}}_x^{\top}\blf D\mathbf{M}_x+d_{N+1}\mathbf{x}_L\widehat{\mathbf{m}}_{x}^{\top}]$. Substituting $\widehat{\mathbf{L}}_x$ and $\widehat{\mathbf{m}}_{x}^{\top}=\mathbf{a}_+^{\o\top}\overline{\mathbf{M}}_{x}+\mathbf{x}_M^{\top}$, we get
\begin{equation}
    \bsym{\Delta}_{12} = \bb E [(\mathbf{L}_x+x_+\overline{\mathbf{L}}_b)^{\top}\blf D\blf M_x
    +d_{N+1}\mathbf{x}_L(\mathbf{a}_+^{\o\top}\overline{\blf M}_x+\mathbf{x}_M^{\top})        ]
\end{equation}
which is in turn composed of the following terms:
%
%
\begin{itemize}
    \item $\bb E [\mathbf{L}_x^{\top}\blf D \blf M_x   ] := \mathbf{L}_t^{\top}\blf D \blf M_t+\sigma^2\blktr(\blf L^{\top}\blf D\blf M,\blf I_N)$ which yields from Lemma \ref{velvet};
    \item $\bb E [x_+\overline{\mathbf{L}}_b^{\top}\blf D\blf M_x] = t_+\overline{\mathbf{L}}_{\mu^\i}^{\top}\blf D\blf M_t$ under the independence consideration;
    \item $\bb E [d_{N+1}\mathbf{x}_L\mathbf{a}_+^{\o\top}\overline{\blf M}_x  ] = d_{N+1}\mathbf{t}_L\bsym{\mu}^{\o\top}\overline{\blf M}_t$ again under independence and where $\mathbf{t}_L = [t_+,\blf 0_L]$;
    \item $d_{N+1}\bb E [\mathbf{x}_L\mathbf{x}_M^{\top}]$. Here, note that $\mathbf{x}_L =[x_+,\blf 0_L]$ and $\mathbf{x}_M = [x_+,\blf 0_M]$. Hence, $\bb E [\mathbf{x}_L\mathbf{x}_M^{\top}]$ equals $\bb E [x_+^2] = t_+^2 + \sigma^2$ in position (1,1) and zero elsewhere. Defining then matrix $\blf T_{LM}\in\mathbb{R}^{L+1\times M+1}$ with  $(t_+^2+\sigma^2)$ in location $(1,1)$ and zero elsewhere, we can write $d_{N+1}\bb E [\mathbf{x}_L\mathbf{x}_M^{\top}] = d_{N+1}\blf T_{LM}$.
\end{itemize}
Combining then these derivations, we get expression \eqref{eq.Delta12} for $\bsym{\Delta}_{12}$.





\smallskip
\noindent\textbf{$\bsym{\Delta}_{21}$.} The third block matrix in \eqref{rad} is $\bsym{\Delta}_{21}:= \bb E[ \mathbf{M}_x^{\top}\blf D\widehat{\mathbf{L}}_x+d_{N+1}\widehat{\mathbf{m}}_{x}\mathbf{x}_L^{\top}]$. It is easy to see that $\bsym{\Delta}_{21} = \bsym{\Delta}_{12}^\top$; hence, \eqref{eq.Delta12}.


\smallskip
\noindent\textbf{$\bsym{\Delta}_{22}$.} The fourth block matrix in \eqref{rad} is $\bsym{\Delta}_{22}:= \bb E [\mathbf{M}_x^{\top}\blf D\mathbf{M}_x+d_{N+1}\widehat{\mathbf{m}}_{x}\widehat{\mathbf{m}}_{x}^{\top}]$. For the first term on the R.H.S. of the latter we have
\begin{equation}\label{eq.dummyD22}
    \bb E [\mathbf{M}_x^{\top}\blf D\mathbf{M}_x] = \mathbf{M}_t^{\top}\blf D\mathbf{M}_t+\sigma^2\blktr({\blf M^{\top}\blf D \blf M},\blf I_N)
\end{equation}
which yields from Lemma \ref{velvet}. Regarding the second term on the R.H.S, we substitute $\widehat{\mathbf{m}}_{x}$ and write it out as
\begin{align}\label{bloatfly}
\begin{split}
&\bb E [d_{N+1}\widehat{\mathbf{m}}_{x}\widehat{\mathbf{m}}_{x}^{\top}]= d_{N+1}\bb E [(\overline{\blf M}_x^{\top}\mathbf{a}_+^{\o}+\mathbf{x}_M)(\overline{\blf M}_x^{\top}\mathbf{a}_+^{\o}+\mathbf{x}_M)^{\top}\\
&=d_{N+1}(\overline{\blf M}_x^{\top}\mathbf{a}_+^{\o}\mathbf{a}_+^{\o\top}\overline{\blf M}_x+\overline{\blf M}_x^{\top}\mathbf{a}_+^{\o}\blf x_M^{\top}+\blf x_M\mathbf{a}_+^{\o\top}\overline{\blf M}_x+\blf x_M\blf x_M^{\top})].
\end{split}
\end{align}
%

We proceed in the same way and elaborate on each terms within the expectation on the R.H.S. of \eqref{bloatfly}; respectively:
\begin{itemize}
    \item $\bb E[\overline{\blf M}_x^{\top}\mathbf{a}_+^{\o}\mathbf{a}_+^{\o\top}\overline{\blf M}_x]=\blktr(\overline{\blf M}^{\top}\blf R^\o\overline{\blf M},(\blf t\blf t^{\top}+\sigma^2\blf I_N))$ which holds from Lemma~\ref{velvet} and where $\blf R^\o=\bsym{\Sigma}^\o+\bsym{\mu}^\o\bsym{\mu}^{\o\top}$;
    \item $\bb E [\overline{\blf M}_x^{\top}\mathbf{a}_+^{\o}\blf x_M^{\top}] = \overline{\blf M}_t^{\top}\bsym{\mu}_+^{\o}\blf t_M^{\top}$;
    \item $\bb E [\blf x_M\mathbf{a}_+^{\o\top}\overline{\blf M}_x] = \blf t_M\bsym{\mu}_+^{\o\top}\overline{\blf M}_t$;
    \item $\bb E [\blf x_M\blf x_M^{\top}] = \text{diag}(t_+^2+\sigma^2,\blf{0}_M))$;
\end{itemize}    
Combining all these terms and \eqref{eq.dummyD22} yields expression \eqref{eq.Delta22} for $\bsym{\Delta}_{22}$.

Next, we focus on the second expectation on the R.H.S. of \eqref{227}: $\boldsymbol{\theta}:= \bb{E}[\bf W_+^{\!\top}\blf D_+\bf t_+]$. Substituting once again $\blf W_+$ [cf. \eqref{eq.filt_compact}], we can write the expectation argument as
%
\begin{align}
\begin{split}
  \blf W_+^{\top}\blf D_+\blf t_+=&\begin{bmatrix}
        \widehat{\mathbf{L}}_x^{\top} &\mathbf{x}_L \\
        \mathbf{M}_x^{\top} & \widehat{\mathbf{m}}_{x}\\
\end{bmatrix}\begin{bmatrix}
        \blf D &\blf 0\\
       \blf 0 & d_{N+1}\\
\end{bmatrix}\begin{bmatrix}
\blf t\\
t_+
\end{bmatrix} \\
&=\begin{bmatrix}
\widehat{\mathbf{L}}_x^{\top}\blf D\blf t +t_+d_{N+1}\blf x_L\\
\mathbf{M}_x^{\top}\blf D\blf t +t_+d_{N+1}\widehat{\mathbf{m}}_{x}
\end{bmatrix}.
\end{split}
\end{align} 
Upon substituting $\widehat{\mathbf{L}}_x$ and $\widehat{\mathbf{m}}_{x}$ and applying the expectation, expression \eqref{eq.theta} for $\bsym \theta $ follows, completing the proof. \hfill\qed
%

\subsection*{Proof of Proposition \ref{Prop 2}}\label{ygai}

\noindent\textbf{Graph $\cal G_+^\i$.} The expected discrete 2-Dirichlet form is
\begin{equation}
    \bb E [S_2(\blf y_+^\i)]=
    \bb E [\blf y^{\i\top}_+(\blf I - \blf A^\i_+)^{\top}(\blf I - \blf A^\i_+)\blf y^{\i}_+]
\end{equation}
which by substituting the adjacency matrix $\blf A^\i_+$ [cf. \eqref{jock}] becomes
\begin{align}\label{eq.dummy0P2}
    \begin{split}
 \bb E [S_2(\blf y_+^\i)]=\bb E \left[\blf y^{\i\top}_+\begin{bmatrix}
 \boldsymbol{\Gamma} & -(\blf I - \blf A)^{\top}\blf b_+^\i \\
 -\blf b_+^{\i\top}(\blf I - \blf A) & \blf b_+^{\i\top}\blf b_+^\i+1
 \end{bmatrix}\blf y^{\i}_+\right]    
    \end{split}
\end{align}
with $\boldsymbol{\Gamma} = (\blf I - \blf A)^{\top}(\blf I - \blf A)$. Substituting further $\blf y^{\i}_+$ [cf. \eqref{eq.outInc}] with $\blf x_+=\blf t_+$ we can write \eqref{eq.dummy0P2} as
\begin{align}\label{ant}
    \begin{split}
      \bb E &[S_2(\blf y^\i_+)]=\blf h^{\i\top}\bb E \bigg[\hat{\mathbf{L}}_t^{\top}\boldsymbol{\Gamma}\hat{\mathbf{L}}_t- \blf t_L \blf b_+^{\i\top}(\blf I - \blf A)\hat{\mathbf{L}}_t\\
      &-\hat{\mathbf{L}}_t^{\top}(\blf I - \blf A)^{\top}\blf b^\i_+\blf t_L^{\top}
      +(\blf b^{\i\top}_+\blf b^\i_++1)\blf t_L\blf t_L^{\top}\bigg]\blf h^\i.
    \end{split}
\end{align}
We now proceed by applying the expectation to each term on the R.H.S. of \eqref{ant}.

%
%
For the first term, we substitute $\hat{\mathbf{L}}_t=(\blf L_t+t_+\overline{\blf L}_{b^\i})$ with $\overline{\blf L}_{b^\i} = \overline{\blf L}_{}|_{x = b^\i} $ and get
\begin{equation}
    \bb E [\hat{\mathbf{L}}_t^{\top}\boldsymbol{\Gamma}\hat{\mathbf{L}}_t] = \bb E[(\blf L_t+t_+\overline{\blf L}_{b^\i})^{\top}\boldsymbol{\Gamma}(\blf L_t+t_+\overline{\blf L}_{b^\i})].
\end{equation}
This is in turn composed of the following four terms:
\begin{itemize}
\item $\bb E[\mathbf{L}_t^{\top}\boldsymbol{\Gamma}\mathbf{L}_t^{\top}]:= \blf L_t^{\top}\bsym{\Gamma}\blf L_t.$ which is unaffected by expectation.

\item $\bb E[t_+\blf L_t^{\top}\bsym{\Gamma}\overline{\blf L}_{b^\i}]:=t_+\blf L_t^{\top}\bsym{\Gamma}\overline{\blf L}_{\bsym{\mu}^\i}$ where we use $\bb E[\blf b^\i]=\bsym{\mu}^\i$.

\item $\bb E[t_+{\blf L}_{\bsym{\mu}^\i}\bsym{\Gamma}\blf L_t]:=t_+\overline{\blf L}_{\bsym{\mu}^\i}^{\top}\bsym{\Gamma}
\blf L_t$

\item $\bb E[t_+^2\overline{\blf L}_{b^\i}^{\top}\bsym{\Gamma}\overline{\blf L}_{b^\i}]:=t_+^2(\overline{\blf L}_{\mu^\i}^{\top}\bsym{\Gamma}\overline{\blf L}_{\mu^\i}+\blktr{(\overline{\blf L}^{\top}\bsym{\Gamma}\overline{\blf L},\bsym{\Sigma}^\i)})$ where we use Lemma \ref{velvet} and $\bb E(\blf b^\i_+\blf b^{\i\top}_+)=\bsym{\Sigma}^\i+\bsym{\mu}^\i\bsym{\mu}^{\i\top}$.
\end{itemize}

The second term in \eqref{ant} $\bb E [\blf t_L \blf b_+^{\i\top}(\blf I - \blf A)\hat{\mathbf{L}}_t]$ is composed of
\begin{itemize}
    \item $\bb E[\blf t_L \blf b_+^{\i\top}\hat{\mathbf{L}}_t]:$ Substituting $\hat{\mathbf{L}}_t=(\blf L_t+t_+\overline{\blf L}_{b^\i})$ we have
  \begin{align}
    \begin{split}
        &\bb{E}[\blf t_L \blf b_+^{\i\top}\hat{\mathbf{L}}_t]=\blf t_L\bb E[\blf b_+^{\i\top}\blf L_t]+t_+\bb E[\blf b_+^{\i\top}\overline{\blf L}_b]
    \end{split}
\end{align}
The term $\blf b_+^{\i\top}\blf L_t$ has expectation $\bsym{\mu}^{\i\top}\blf L_t$. We also have $\blf b_+^{\i\top}\overline{\blf L}_b=\begin{bmatrix}\blf 0 & \blf b_+^{\i\top}\blf A\blf b_+^\i & \ldots & \blf b_+^{\i\top}\blf A^{L-1}\blf b_+^\i \end{bmatrix}$. Using Lemma~\ref{velvet}, we have $\bb E [\blf b_+^{\i\top}\overline{\blf L}_b]=\bsym{\mu}^{\i\top}\overline{\blf L}_{\mu^\i}+\blktr(\overline{\blf L},\bsym{\Sigma}^\i)$. Combining, we get
\begin{align}\label{jj}
\begin{split}
 &\bb{E}[\blf t_L \blf b_+^{\i\top}\hat{\mathbf{L}}_t]=\blf t_L(\bsym{\mu}_+^{\i\top}\blf L_t+t_+(\bsym{\mu}^{\i\top}\overline{\blf L}_{\mu^\i}+\blktr(\overline{\blf L},\bsym{\Sigma}^\i))).
\end{split}
\end{align}
\item $\bb{E}[\blf t_L \blf b_+^{\i\top}\blf A\hat{\mathbf{L}}_t]=\blf t_L\bb E[\blf b_+^{\i\top}\blf A\blf L_t]+t_+\bb E[\blf b_+^{\i\top}\blf A\overline{\blf L}_b]$, which by following similar arguments yields
\begin{align}\label{ii}
    \begin{split}
        &\bb{E}[\blf t_L \blf b_+^{\i\top}\blf A\hat{\mathbf{L}}_t]=\blf t_L(\bsym{\mu}^{\i\top}\blf A\blf L_t+t_+(\bsym{\mu}^{\i\top}\blf A\overline{\blf L}_{\mu^\i}+\blktr(\blf A\overline{\blf L},\bsym{\Sigma}^\i))).
    \end{split}
\end{align}
\end{itemize}

Combining \eqref{jj} and \eqref{ii}, we get
\begin{align}\label{the}
    \begin{split}
 & \bb E[\blf t_L \blf b_+^{\i\top}(\blf I - \blf A)\hat{\mathbf{L}}_t]=\blf t_L(\bsym{\mu}^{\i\top}\blf L_t+t_+(\bsym{\mu}^{\i\top}\overline{\blf L}_{\mu^\i}+\blktr(\overline{\blf L},\bsym{\Sigma}^\i)))\\
&-\blf t_L(\bsym{\mu}^{\i\top}\blf A\blf L_t+t_+(\bsym{\mu}^{\i\top}\blf A\overline{\blf L}_{\mu^\i}+\blktr(\blf A\overline{\blf L},\bsym{\Sigma}^\i))).     
    \end{split}
\end{align}

The third term is the transpose of the second one, i.e.,
\begin{align}\label{nn}
    \begin{split}
        &\bb E[\hat{\mathbf{L}}_t^{\top}(\blf I - \blf A)^{\top}\blf b^\i_+\blf t_L^{\top}]= \bb E[\blf t_L \blf b_+^{\i\top}(\blf I - \blf A)\hat{\mathbf{L}}_t]^{\top}
    \end{split}
\end{align}

The final term is
\begin{align}
    \bb E[(\blf b^{\i\top}_+\blf b^\i_++1)\blf t_L\blf t_L^{\top}]=\big(\sum_{n=1}^N\bb E[[\blf b_+^\i]_n^2]+1\big)\textnormal{diag}(t_+^2,\blf 0).
\end{align}where we used $\blf t_L\blf t_L^{\top}=\textnormal{diag}(t_+^2,\blf 0)$. Given that $\bb E[[\blf b_+^\i]_n^2]=w_n^2p_n$, we can write this as
\begin{align}\label{hell}
    \bb E[(\blf b^{\i\top}_+\blf b^\i_++1)\blf t_L\blf t_L^{\top}]=(\blf w^{\top}\blf p^\i+1)\textnormal{diag}(t_+^2,\blf 0).
\end{align}

Combining these together we get expression \eqref{eq.psii} for $\bsym{\Psi}^\i$.

\smallskip
\noindent\textbf{Graph $\cal G_+^\o$.} By substituting $\blf A^\o_+$ [cf. \eqref{jock}] and $\blf y^{\o}_+=[\blf M_t\blf h^\o,\widehat{\mathbf{m}}_{t}^{\top}\blf h^\o]^{\top}$ in $\blf y^{\o\top}_+(\blf I - \blf A^\o_+)^{\top}(\blf I - \blf A^\o_+)\blf y^{\o}_+$ the expected 2-Dirichlet form is
\begin{align}
\begin{split}
  &\bb E[S_2(\blf y^\o_+)] \!=\! \bb E \bigg[\!\! \begin{bmatrix}
 \blf h^{\o\top}\blf M_t^{\top} &\!\! \blf h^{\o\top}\widehat{\mathbf{m}}_{t}
 \end{bmatrix}\!\!\begin{bmatrix}
 \bsym{\Gamma}+\blf a^{\o}_+\blf a^{\o\top}_+ &\!\! -\blf a_+^{\o} \\
 -\blf a_+^{\o\top} &\!\! 1
 \end{bmatrix}\begin{bmatrix}
 \blf M_t\blf h^\o \\ \widehat{\mathbf{m}}_{t}^{\top}\blf h^\o
 \end{bmatrix} \!\!\bigg]. 
 \end{split}
 \end{align}
Utilizing $\widehat{\mathbf{m}}_{t}=\overline{\blf M}_t^{\top}\blf a^{\o}+\blf t_M$, we get
 \begin{align}\label{jgraham}
 \begin{split}
  &\bb E [S_2(\blf y^\o_+)]=\blf h^{\o\top}\bb E\bigg[\blf M_t^{\top}(\bsym{\Gamma}+\blf a^{\o}_+\blf a^{\o\top}_+)\blf M_t\\
  &-\blf M_t^{\top}\blf a^{\o}_+(\blf a^{\o\top}_+\overline{\blf M}_t+\blf t_M^{\top})-(\overline{\blf M}_t^{\top}\blf a^{\o}+\blf t_M)\blf a^{\o\top}_+\blf M_t\\
  &+(\overline{\blf M}_t^{\top}\blf a^{\o}+\blf t_M)(\overline{\blf M}_t^{\top}\blf a^{\o}+\blf t_M)^{\top}\bigg]\blf h^\o.
  \end{split}
\end{align}
\par Equation \eqref{jgraham} comprises the following four terms:
\begin{itemize}
    \item $\bb E[\blf M_t^{\top}(\bsym{\Gamma}+\blf a^{\o}_+\blf a^{\o\top}_+)\blf M_t] = \blf M_t^{\top}(\bsym{\Gamma}+\blf R^\o)\blf M_t$ since $\blf a^{\o}_+$ is random here and $\blf R^\o=\bb E[\blf a_+^\i\blf a_+^{\i\top}]$.
    \item $\bb E[\blf M_t^{\top}\blf a^{\o}_+(\blf a^{\o\top}_+\overline{\blf M}_t+\blf t_M^{\top})]=\bb E[\blf M_t^{\top}\blf a^{\o}_+\blf a^{\o\top}_+\overline{\blf M}_t]+\bb E[\blf M_t^{\top}\blf a^{\o}_+\blf t_M^{\top}] = \blf M_t^{\top}\blf R^\o\overline{\blf M}_t +\blf M_t^{\top}\bsym {\mu}^{\o}_+\blf t_M^{\top}$. 
    \item The third term is the transpose of the above, hence it has the expectation $\overline{\blf M}_t^{\top}\blf R^\o\blf M_t+\blf t_M\bsym{\mu}^{\o\top}\blf M_t$.
    \item $\bb E[(\overline{\blf M}_t^{\top}\blf a^{\o}+\blf t_M)(\overline{\blf M}_t^{\top}\blf a^{\o}+\blf t_M)^{\top}]=\bb E[\overline{\blf M}_t^{\top}\blf a^{\o}\blf a^{\o\top}\overline{\blf M}_t]+\bb E[\overline{\blf M}_t^{\top}\blf a^{\o}\blf t_M^{\top}]+\bb E[\blf t_M\blf a^{\o\top}\overline{\blf M}_t]+\bb E[\blf t_M\blf t_M^{\top}] = \overline{\blf M}_t^{\top}\blf R^\o\overline{\blf M}_t+\overline{\blf M}_t^{\top}\bsym {\mu}^{\o}\blf t_M^{\top}+\blf t_M\bsym {\mu}^{\o\top}\overline{\blf M}_t+\blf t_M\blf t_M^{\top}$ by operating term-wise. 
%
\end{itemize}
Combining these together, we get expression \eqref{eq.psio} for $\bsym{\Psi}^\o$; hence, completing the proof.\hfill\qed
\end{document}